\documentclass[reprint,amsmath,amssymb,aps,floatfix,pre,superscriptaddress]{revtex4-1}

\usepackage{graphicx}
\usepackage{dcolumn}
\usepackage{bm}

\begin{document}

\preprint{}

\title{Ground state structures of superparamagnetic 2D dusty plasma crystals}

\author{Peter~Hartmann}
\affiliation{Research Institute for Solid State Physics and Optics of the Hungarian Academy of Sciences, H-1525 Budapest, P.O. Box 49, Hungary}
\affiliation{Department of Physics, 
Boston College, Chestnut Hill, MA 02467}
\author{Marlene~Rosenberg}
\affiliation{Department of Electrical and Computer Engineering, University 
of California San Diego, La Jolla CA 92093}
\author{Gabor~J.~Kalman}
\affiliation{Department of Physics, 
Boston College, Chestnut Hill, MA 02467}
\author{Zolt\'an~Donk\'o}
\affiliation{Research Institute for Solid State Physics and Optics of the Hungarian Academy of Sciences, H-1525 Budapest, P.O. Box 49, Hungary}
\affiliation{Department of Physics, 
Boston College, Chestnut Hill, MA 02467}

\begin{abstract}
Ground state structures of finite, cylindrically confined two-dimensional Yukawa systems composed of charged superparamagnetic dust grains in an external magnetic field are investigated numerically, using molecular dynamic simulations and lattice summation methods. The ground state configuration of the system is identified using, as an approximation, the experimentally obtained shape of the horizontal confinement potential in a classical single layer dusty plasma experiment with non-magnetic grains. Results are presented for the dependence of the number density and lattice parameters of the dust layer on (1) the ratio of the magnetic dipole-dipole force to electrostatic force between the grains and (2) the orientation of the grain magnetic moment with respect to the layer.   
\end{abstract}

\pacs{}

\maketitle

\section{Introduction}

Plasma crystals composed of dust grains that are superparamagnetic, where each grain can acquire a strong magnetic dipole moment in a  magnetic field, are expected to lead to  new possibilities in dusty plasma research \cite{Samsonov03}. While the electrostatic interaction between the negatively charged grains is repulsive and isotropic, the magnetic dipole-dipole interaction is in general anisotropic and can be attractive or repulsive as a function of orientation of the magnetic dipole moments \cite{Yaro04}. 

The use of superparamagnetic grains could enable the magnetic tuning of plasma crystal structures, similar to what has been considered for superparamagnetic colloidal crystals (e.g. \cite{Xu01, Xu02, Frolt03, Pu08}). Very recently, colloidal suspensions of sub-micron sized ($\approx 100$~nm) polyacrylate capped superparamagnetic magnetite (${\rm Fe_3O_4}$) particles were successfully used to produce colloidal photonic crystals with magnetically tunable stop bands covering the visible spectrum \cite{coll1,coll2,coll3}. The superparamagnetic colloids form chain-like structures along an external magnetic field with regular inter-particle spacing, enabling the diffraction of visible light. The tuning of the diffraction wavelength was accomplished by varying the inter-particle spacing. In turn this was done by varying the magnetic field that alters the strength of the magnetic dipole-dipole interaction, which balances the repulsive electrostatic interaction between the charged colloids \cite{coll1, coll2}. Single layer experiments with superparamagnetic particles on the water-air interface have demonstrated the advantages of the tunable inter-particle interaction in the studies of fundamental collective phenomena, like the solid-liquid phase transition \cite{Keim_melt,Keim_quench}. Coagulation of charged, charged-magnetic, and magnetic dust aggregates formed from a ferrous material in various environments was studied in \cite{Perry10}, showing that the dipole-dipole interaction can affect the orientation and structural formation of aggregates as they collide and stick.
 
While colloidal crystals typically have inter-particle spacings in the sub-micron regime, the spacing between dust grains in plasma crystals is typically larger, on the order of 100 $\mu$m, which is in the range of terahertz (THz) wavelengths. We investigate the possibility of using superparamagnetic particles in the larger micrometer size range in a dusty plasma monolayer in a magnetic field, with the aim of producing a tunable two-dimensional (2D) lattice structure with spacings that correspond to the THz regime \cite{THz}. The tuning is accomplished by varying the angle the magnetic field subtends with the plane of grains. If such structures can be produced with the grains occupying a large volume fraction of dust grains (see \cite{THz}), they may have photonic applications in the THz frequency range which is currently of great interest owing to potential applications in spectroscopy, imaging, etc. \cite{Drago2}. 

The paper is organized as follows. Section II presents the model system, which is a confined 2D layer of charged superparamagnetic grains in a plasma, placed in an external magnetic field. Section III presents the results of MD simulations of the ground state structures of a finite 2D system in the crystalline solid phase as the relative strength of the electrostatic to magnetic dipole-dipole interaction and the direction of the grains' magnetic moment with respect to the layer plane are varied. Section IV presents lattice summation calculations of the corresponding infinite 2D lattice limit of this system. A discussion of possible experimental parameters is given in Section V, and a brief summary is given in section VI.

\section{Model system}

\begin{figure}[htb]
\includegraphics[width=1.0\columnwidth]{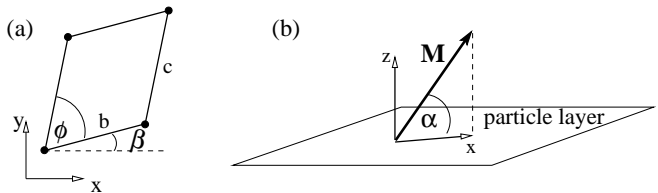}
\caption{\label{fig:1} 
Geometry of the model system. (a) The lattice lies in the $x-y$ plane and its structure is characterized by the lattice spacing $b$, the rhombic angle $\phi$ and the aspect ratio $\nu = c/b$. (b) The magnetic moment ${\bf M}$ of each grain lies in the $x-z$ plane at an angle $\alpha$ to the $x$-axis. The principal lattice axes subtends an angle $\beta$ with the projection of the magnetic moment.}
\end{figure}

The model system comprises a 2D lattice of superparamagnetic dust grains immersed in a plasma in a constant, homogeneous external magnetic field ${\bf B}$. Each grain acquires an electric charge $q$ due to plasma collection, and a magnetic dipole moment ${\bf M}$, which is induced by the external magnetic field and therefore lies in the direction of ${\bf B}$. The lattice lies in the $x-y$ plane with an unspecified orientation of its principal axes. The lattice structure is characterized by the lattice spacing $b$, the rhombic angle $\phi$ and the aspect ratio $\nu = c/b\ge1$, and the direction of its principal axes with respect to the projection of the magnetic field onto the plane, as shown in Fig.~\ref{fig:1}. 

The grains interact via an electrostatic screened Coulomb (Debye-H\"uckel or Yukawa) force and by the induced magnetic dipole-dipole force. The electrostatic interaction energy between two grains with charge $q$ separated by a distance $r$ is 
\begin{equation}
U_E = q^2/(4\pi\varepsilon_0) {\rm exp}(-r/\lambda_D) /r, 
\end{equation}
where $\lambda_D$ is the plasma Debye screening length, yielding a repulsive force  
\begin{equation}\label{eq:1}
{\bf F}_E(r) =  \frac{1}{4\pi\varepsilon_0}\frac{q^2}{r^2} \left(1 + \frac{r}{\lambda_D}\right) {\rm exp} \left(-\frac{r}{\lambda_D}\right)\hat{\bf r},
\end{equation}
where $\hat{\bf r}$ is a unit vector in the direction of ${\bf r}$, which is the vector connecting the two particles. The magnetic dipole-dipole force between two grains, ${\bf F}_M$ can be repulsive or attractive depending on the relative positions and orientations of the grains. Since it is assumed that the magnetic dipole moments of all the grains are parallel and have the same magnitude, the interaction energy of the two magnetic dipoles is given by 
\begin{equation}\label{eq:2}
U_M = \frac{\mu_0}{4\pi}\left[\frac{M^2}{r^3} - \frac{3 ({\bf M} \cdot {\bf r})^2}{r^5}\right].
\end{equation}
The magnetic dipole-dipole force between the two grains is
\begin{equation}\label{eq:3}
{\bf F}_M = \frac{\mu_0}{4\pi}\frac{3M^2}{r^4} [- \hat{\bf r} (5 {\rm cos}^2 \,\theta -1) + 2 \hat{\bf m}~{\rm cos} \theta],
\end{equation}
where $\hat{{\bf r}}$ and $\hat{{\bf m}}$ are unit vectors in the direction of ${\bf r}$ and ${\bf M}$, respectively, and $\theta$ is the angle between $\hat{{\bf r}}$ and $\hat{{\bf m}}$. In the following we choose without loss of generality our coordinate system such that ${\bf M}$ is in the $x-z$ plane and is oriented at an angle $\alpha$ with respect to the $x$-axis (see Fig.~\ref{fig:1}).  
 
\begin{figure}[htb]
\includegraphics[width=0.9\columnwidth]{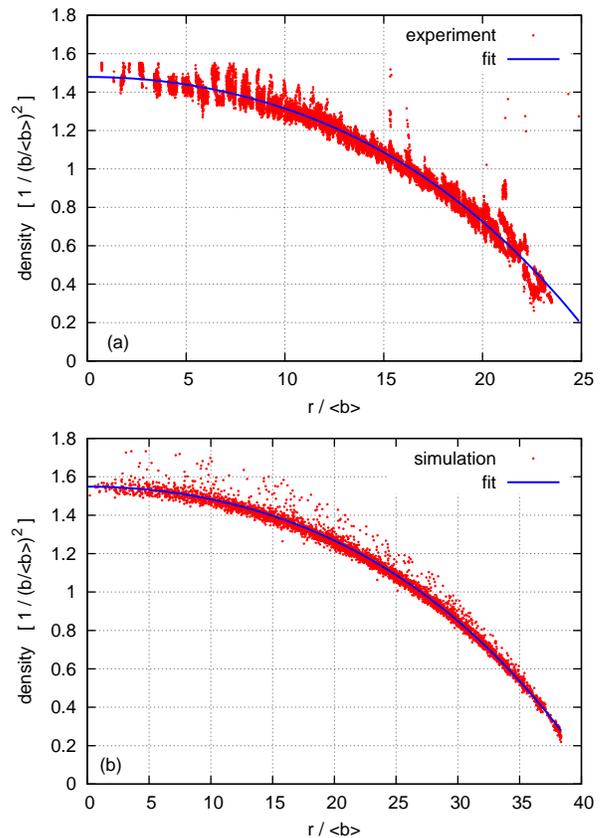}
\caption{\label{fig:2} 
(color online) Experimental (a) and MD simulation (b) results for the radial dust density distribution in 2D layer of non-magnetic dust. Distances are normalized to the observed average lattice spacing $\langle b\rangle$. The total particle numbers are $\sim 3000$ in the experiment and 5000 in the simulation.}
\end{figure}
 
In a typical 2D dusty plasma laboratory experiment, the dust grains are confined by an electrostatic potential. In order to approximate experimental conditions in our simulations, we considered a more accurate representation of the radial dependence of the horizontal confinement potential beyond the usual quadratic approximation. This was done by performing an experiment using non-magnetic melamine-formaldehyde (MF) particles with the aim of measuring the radial density profile of the single layer dust cloud. Without going in the details, the experiments used 4.36 $\mu$m diameter MF spheres, in a 1 Pa argon gas discharge driven by 5 W of RF power at 13.56 MHz. A layer of $\sim 3000$ MF spheres was created over the 18 cm diameter lower powered electrode. Particle detection was performed using 650~nm wavelength laser illumination from the side and a 1.4 MPixel CCD camera from the top. Sub-pixel resolution was achieved using the center-of-mass method discussed in detail in \cite{Feng_piv}. Calculating the density of the dust layer by averaging over the nearest-neighbor distances, the experimental density profile was approximated by the functional form:
\begin{equation}\label{eq:4}
n(\bar{r})  \approx n_4\bar{r}^4+n_2\bar{r}^2+n_0,
\end{equation}
where $\bar{r} = r/\langle b \rangle$ and $\langle b \rangle$ is the average lattice spacing, as shown in Fig.~\ref{fig:2}(a). The corresponding molecular dynamics (MD) simulation results using $N=5000$ particles and an average Coulomb coupling parameter $\Gamma \approx 1000$ is displayed in Fig.~\ref{fig:2}(b). In contrast to infinite, homogeneous systems, the Coulomb coupling parameter does not fully characterizes the entire particle ensemble as it depends on the particle density, which has a strong radial profile in our confined system. Details of the simulation model can be found in the next section.

The experimental dust density profile could be best reproduced assuming a horizontal confinement potential of the form 
\begin{equation}\label{eq:5}
V(r) = V_4r^4+V_2r^2,
\end{equation} 
with $V_4=5\cdot 10^{-7}$ and $V_2=0.004$. Here and in the following, we use distances normalized to the Debye screening length $\lambda_D$, kept constant for all simulations.

Note that compared to a confinement potential with a simple quadratic dependence on $r$, (\ref{eq:5}) this simulation leads to a more homogeneous distribution in the center of the cloud, with about 10 to 20 \% lower density and has resulted in $\langle b \rangle \approx \lambda_D$ and a dimensionless screening parameter $\kappa=(\lambda_D\sqrt{\pi n})^{-1} \approx 0.53$ in the central region.

\section{Molecular Dynamics Simulations}

The molecular dynamics (MD) simulations are based on a standard method described in e.g. \cite{MD}. We consider a 2D layer particle ensemble of 5000 particles. Pair interactions (forces) are evaluated in every time-step for each pair of particles. Time integration is performed using the velocity-Verlet scheme. Particles are released from random positions, a slow velocity back-scaling thermostat is applied until the system reached an average Coulomb coupling parameter of 1000. Simulations were run for about 1000 plasma oscillation cycles without further thermostation assuming that a near to ground state configuration could develop during this time. 

\begin{figure}[htb]
\includegraphics[width=0.9\columnwidth]{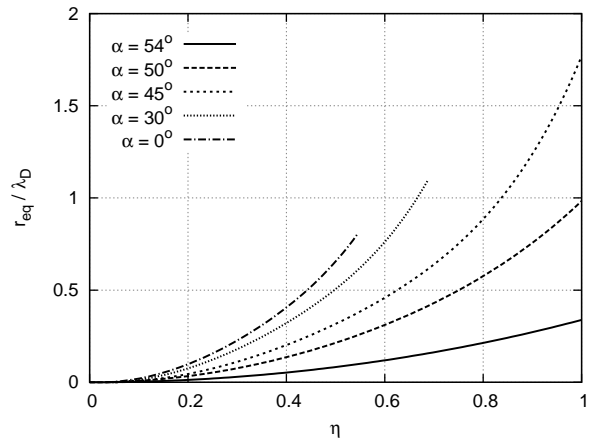}
\caption{\label{fig:3} 
Equilibrium distance $r_{eq}({\bf M}, \alpha)$ for $\alpha < \alpha_\text{th}$ along the $x$-direction versus $\eta$. For strong magnetic interactions (large $\eta$) the attraction dominates at all distances, thus no equilibrium distance can be found, as indicated by the discontinuation of the lines for $\alpha = 0^\text{o}$ and $30^\text{o}$.} 
\end{figure}

For the simulations and the presentation of our results we use the following reduced quantities: $\lambda_D = 1$ (length unit), $b$ is the lattice spacing in units of $\lambda_D$, $q=1$ (charge unit) is the dust grain charge, and $\eta = \sqrt{\mu_0\varepsilon_0}M/q \lambda_D$ is a measure of the relative strength of the magnetic dipole-dipole interaction to the electrostatic interaction. The layer structure is further characterized by the bulk density $n$, both $b$ and $n$ being an average over the central region of the layer, along with the rhombic angle $\phi$ and aspect ratio $\nu$. 

\begin{figure}[htb]
\includegraphics[width=0.9\columnwidth]{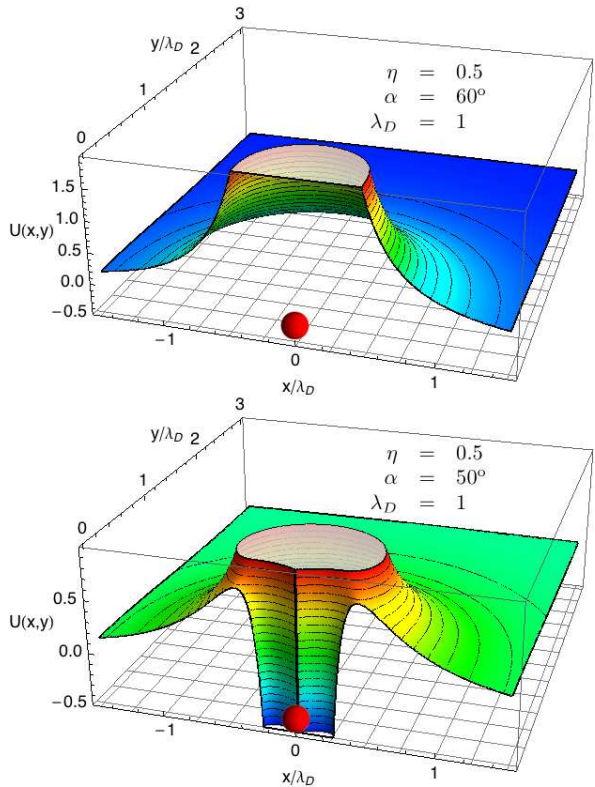}
\caption{\label{fig:field1}  
(color online) Total pair potential energy $U({\bf r}) = U_E({\bf r}) + U_M({\bf r})$ around a single particle situated at ${\bf r}=(0,0)$ at a relative strength of the magnetic interaction to the electrostatic interaction $\eta=0.5$: (a) $\alpha = 60^\text{o} > \alpha_\text{th}$ and (b) $\alpha = 50^\text{o} < \alpha_\text{th}$. The potential energy surface is cut at $y=0$ to display the variation of the interaction energy along the $x$-axis. The central particle is shown.}  
\end{figure}

Setting ${\bf F}_M({\bf r},{\bf M},\alpha) + {\bf F}_E({\bf r}) = 0$, yields the equilibrium distance $r_\text{eq}({\bf M},\alpha)$ where the electrostatic and magnetic forces balance. The effect of the magnetic field is the strongest when ${\bf r}$ is purely in the $x$-direction, in this case Eq. (\ref{eq:3}) yields   
\begin{equation}\label{eq:6}
F_M(x,M,\alpha) = - \frac{\mu_0}{4\pi}\frac{3M^2}{x^4} \left[3({\rm cos}^2\alpha) \,-1\right].
\end{equation}  
In this case, there is a threshold angle, $\alpha_\text{th}=\cos^{-1}(1/\sqrt{3}) \approx 54.74^\text{o}$, below which the attractive interaction due to the magnetic dipole-dipole force can overcome the repulsive electrostatic interaction for certain values of $\eta$, and agglomeration can set in. The variation of $r_\text{eq}$ with $\eta$ is shown in Fig.~\ref{fig:3} for several values of $\alpha$ that are below the threshold angle. As expected, $r_\text{eq}$ increases as $\eta$ increases, with the largest increase for small $\alpha$ since the magnitude of the attractive interaction gets larger as $\alpha$ gets smaller. Thus the grains could agglomerate at progressively smaller values of $\eta$ as $\alpha$ decreases. Furthermore at large enough $\eta$ values the magnetic attraction fully dominates over the electrostatic repulsion, thus an equilibrium distance can not be found at all. This trend will also be apparent in the following discussions of the MD simulation results on the variation of the structure of the lattice under variation of $\eta$ and $\alpha$. 

An illustration of the effect of the competing magnetic and electrostatic interactions is shown in Fig.~\ref{fig:field1}, displaying the total pair potential energy $U({\bf r}) = U_E({\bf r}) + U_M({\bf r})$ of a single particle for selected $\alpha$ angles above and below the threshold value and $\eta = 0.5$. The interaction is repulsive in all directions for $\alpha = 60^\text{o}$; for $\alpha = 50^\text{o}$ around $x=0$ an attractive region develops in a narrow angle around the $\pm x$ direction, separated by a potential barrier from the outside, as it can be seen from this cross-section at $y=0$ (front face). Particles with high enough energy in the tail of the thermal distribution can overcome the potential barrier and result in particle agglomeration after a long enough time.

\begin{figure}[htb]
\includegraphics[width=0.9\columnwidth]{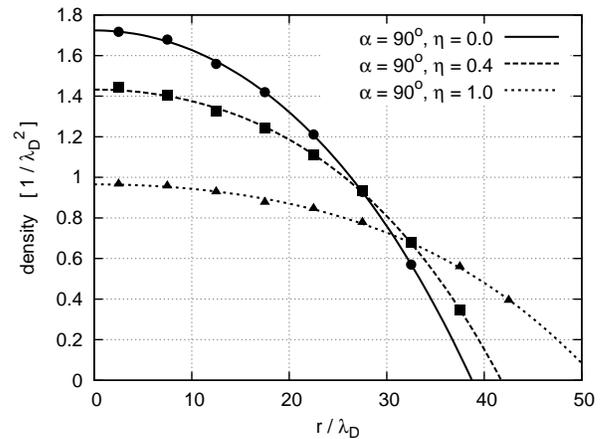}
\caption{\label{fig:4}  
Radial density distribution in the dust layer (in units of $1/\lambda_D^2$) as a function of $r$ (in units of $\lambda_D$), for $\alpha = 90^\text{o}$ and several values of $\eta$.}  
\end{figure}

\begin{figure}[htb]
\includegraphics[width=0.9\columnwidth]{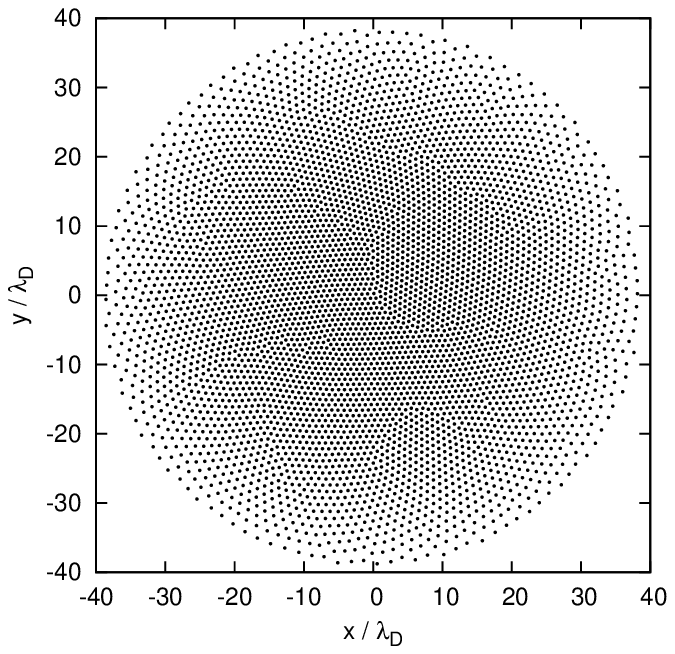}
\caption{\label{fig:5}  
Snapshot of layer for $\alpha = 90^\text{o}$ and $\eta = 0.1$}  
\end{figure}

Turning now to the lattice structure, first consider the case when there is no magnetic field, so that there are no induced magnetic moments ($\eta = 0$). The underlying hexagonal structure is due to the isotropic repulsive electrostatic interaction and is characterized by $\phi = 60^\text{o}$ and $\nu = 1$. Due to the boundary condition imposed by the cylindrical symmetry of the confinement and to the fact that a perfect hexagonal configuration cannot form in a system with density gradinet, lattice frustrations result in slight fragmentation of the ground state structure. Next, consider the case there is an external magnetic field perpendicular to the layer, $\alpha = 90^\text{o}$. The lattice structure remains hexagonal, since both the magnetic dipole-dipole and electrostatic interactions are repulsive and isotropic. As expected, the density decreases as $\eta$ increases (as $q=1$ and $\lambda_D=1$ are kept constant), that is, the average lattice spacing increases owing to the increased repulsive force, as can be seen in the density profile results in Fig.~\ref{fig:4}. Figure~\ref{fig:5} shows a snapshot of the system for $\alpha = 90^\text{o}$ and $\eta = 0.1$. 

\begin{figure}[htb]
\includegraphics[width=0.9\columnwidth]{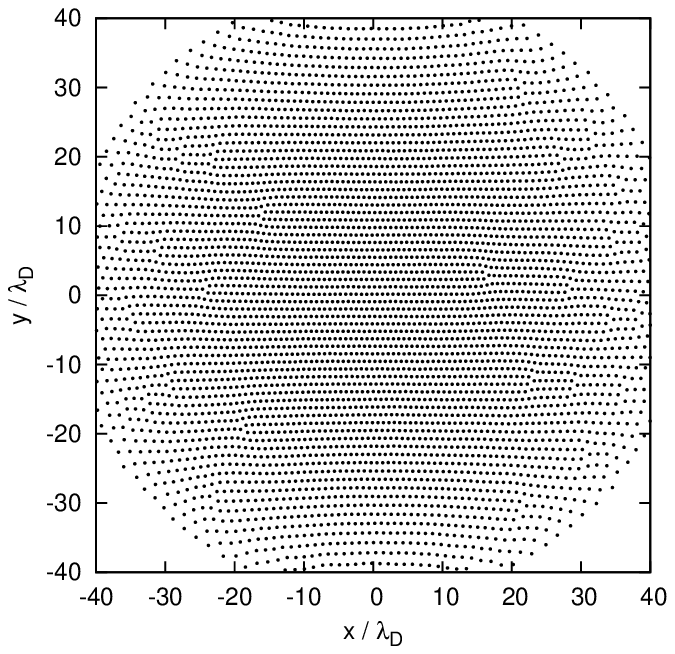}
\caption{\label{fig:6}  
Snapshot of layer for $\alpha = 60^\text{o}$ and $\eta = 0.8$}  
\end{figure}

\begin{figure}[htb]
\includegraphics[width=0.9\columnwidth]{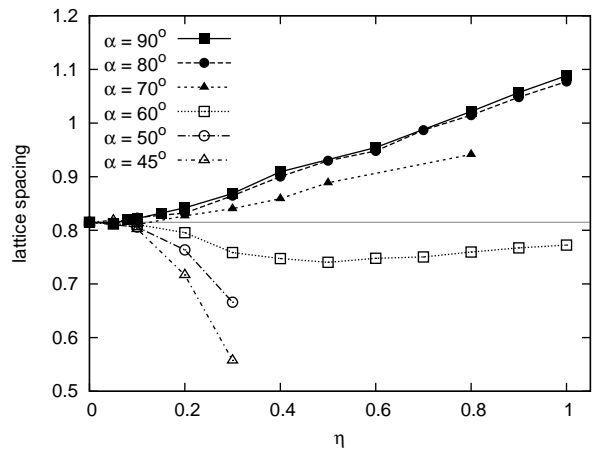}
\caption{\label{fig:7}  
Average lattice spacing $b$ (in units of $\lambda_D$) versus $\eta$, for various 
values of angle $\alpha$.} 
\end{figure}

\begin{figure}[htb]
\includegraphics[width=0.9\columnwidth]{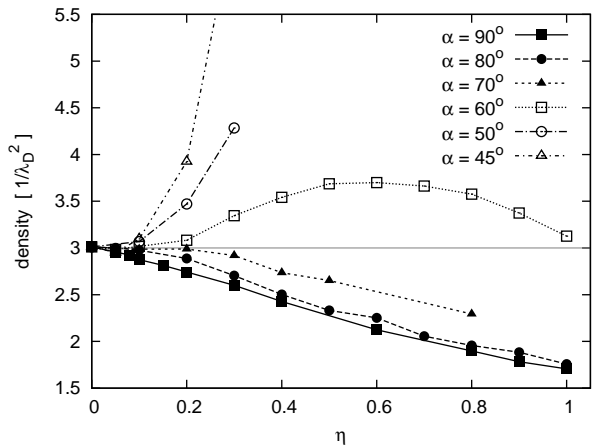}
\caption{\label{fig:8}  
Dust particle central density (in units of $1/\lambda_D^2$) versus $\eta$, for various values of angle $\alpha$.} 
\end{figure}

\begin{figure}[htb]
\includegraphics[width=0.9\columnwidth]{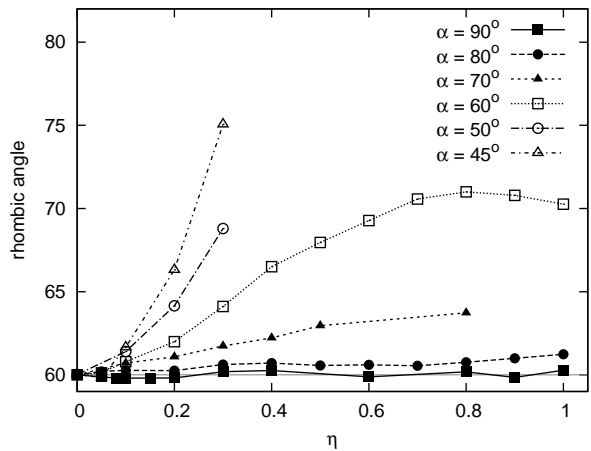}
\caption{\label{fig:9}  
Rhombic angle $\phi$ versus $\eta$, for various values of angle $\alpha$.} 
\end{figure}

\begin{figure}[htb]
\includegraphics[width=0.9\columnwidth]{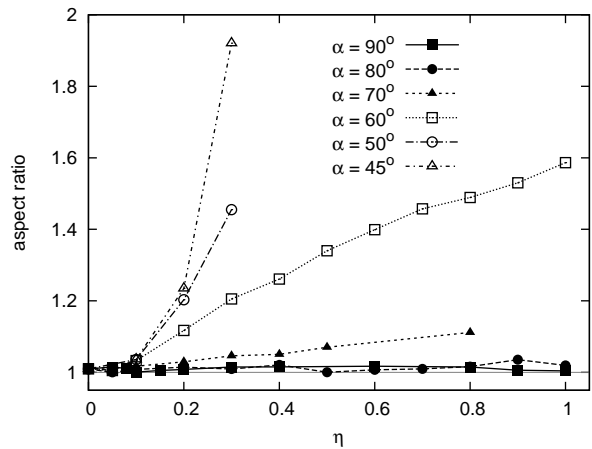}
\caption{\label{fig:10}  
Aspect ratio $\nu$ versus $\eta$, for various values of angle $\alpha$.} 
\end{figure}

Consider now the more interesting cases when the direction of the induced magnetic moments is tilted with respect to the dust layer ($\alpha < 90^\text{o}$). The shortest lattice distance forms along the $x$-axis, thus the lattice forms with $\beta=0$ (see Fig~\ref{fig:1}), as might be expected since the magnetic repulsion weakens or eventually turns purely attractive in that direction (depending on the value of $\alpha$). Thus the system appears to align along that direction. Figure~\ref{fig:6} shows a snapshot of the system for $\alpha = 60^\text{o}$ and $\eta = 0.8$, where a crystal structure without domain fragmentation is formed, showing that the ordering effect arising from the magnetic enhancement of the interaction overcomes the frustration induced by the boundary condition. The variation of the lattice spacing $b$, the bulk density $n$, the rhombic angle $\phi$ and the aspect ratio $\nu$ are shown in Figs.~\ref{fig:7}-\ref{fig:10}, respectively, as a function of $\eta$ for various values of $\alpha$. For the large angles $\alpha \gtrsim 70^\text{o}$, the lattice spacing increases and the density decreases as $\eta$ increases, because the magnetic dipole-dipole interaction is repulsive and its anisotropy is not strong. For this range of angles $\alpha$, both the rhombic angle and aspect ratio of the lattice increase somewhat as $\alpha$ decreases, as the magnetic interaction becomes less repulsive. For the small angles, $\alpha = 45^\text{o}$ and $50^\text{o}$, the lattice spacing decreases as $\eta$ increases, owing presumably to the dominance of the attractive magnetic interaction, which significantly weakens the electrostatic repulsion. At this low $\alpha$ angles the system becomes unstable against aggregation, in the sense discussed above, at intermediate $\eta$ values. In the true ($T=0$) ground state, low $\alpha$ configurations are stable as long as $b>r_{eq}$, however our simulations are run at very low, but finite temperatures, where agglomeration can start (causing the simulation to stop) due to thermal energy fluctuations at long enough times. The $\eta$ values at which this occurs ($\eta \approx 0.3$ ) are for this particular set of simulation parameters, simulation time in particular. During this time the system reaches only kinetically stable states, and not a thermodynamical equilibrium state.

\begin{figure}[htb]
\includegraphics[width=0.9\columnwidth]{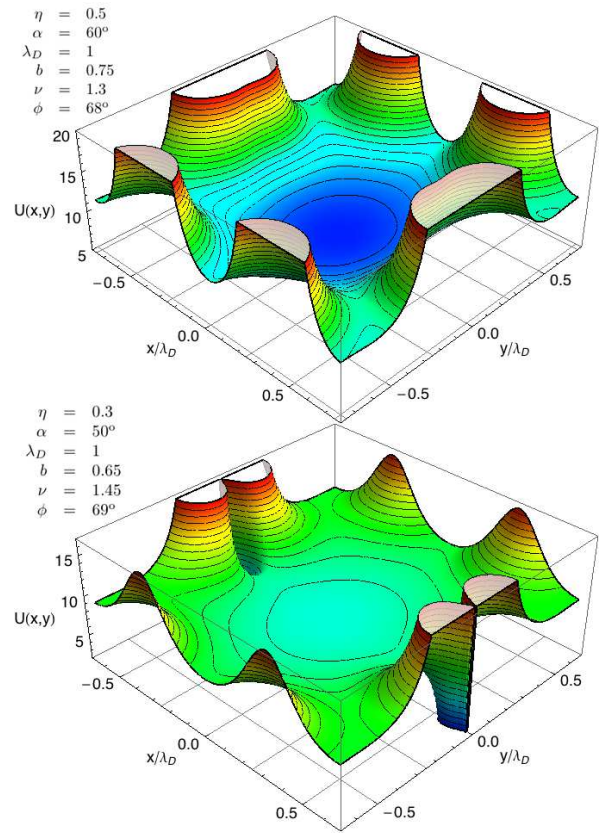}
\caption{\label{fig:field2}  
(color online) Total potential energy $U({\bf r}) = U_E({\bf r}) + U_M({\bf r})$ of the lattice experienced by  a test particle situated at ${\bf r}=(0,0)$. Lattice parameters are taken from the MD simulations for (a) an $\alpha>\alpha_\text{th}$  and (b) an $\alpha<\alpha_\text{th}$ configuration. The area plotted is restricted to $y \geq 0$. Two particles are situated at the front left and right corners, two other particles are shown at their lattice positions with their full symmetric (magnetic) dipole + (electrostatic) monopole fields.}  
\end{figure}

This is illustrated in Fig.~\ref{fig:field2}, where the total potential due to the lattice particles, as experienced by a particle in the center is shown. In the large $\alpha>\alpha_\text{th}$ case (a) the potential energy $U({\bf r})$ surface exhibits a deep, well confined potential minimum. This can be contrasted with the case of a selected low $\alpha<\alpha_\text{th}$ value (b) where a minimum enclosed by a low potential barrier is formed along the $\pm x$ directions around the vacant particle position at ${\bf r}=(0,0)$.

While the central density tends to increase somewhat with $\eta$, both the rhombic angle and aspect ratio increase rapidly, tending toward a rectangular configuration. 

At the intermediate angle $\alpha = 60^\text{o}$, there appears to be non-monotonic characteristics of some of the lattice parameters. When $\eta$ is small, the trends follow those described previously for small $\alpha$, with $b$ decreasing, and $\phi$ and $\nu$ increasing, as $\eta$ increases. However, at larger $\eta \gtrsim 0.4-0.5$, the lattice spacing $b$ begins to somewhat increase, although still remaining below its value at $\alpha = 90^\text{o}$. Meanwhile the central density decreases significantly, which may indicate that the overall magnetic repulsion starts to overcome the complex effect of the force anisotropy. This non-monotonic behavior may be consistent with the trends pointed out in Fig.~\ref{fig:4} in \cite{Samsonov03}, for intermediate magnetic field values, where it was found that for small inter-grain distances, the total force $F_E + F_M$ from eqs. (\ref{eq:1}) and (\ref{eq:3}) was attractive, while at intermediate distances the total force was repulsive.  

Overall, it can be seen from Figs.~\ref{fig:7}-\ref{fig:10} that for these parameters, it may be possible to tune the lattice spacing and structure by changing $\eta$ and $\alpha$. The lattice spacing could be tuned by a factor of about 2, with a corresponding factor of $\sim 2$ in the particle density and changing together with other lattice parameters. The lattice structure could be tuned from triangular (hexagonal) to almost rectangular; depending on $\eta$ and $\alpha$, the rhombic angle can vary between 60 and 80 degrees, and the aspect ratio between 1 and 2.

\section{Infinite Lattice}

We have investigated the ground state energy at $T=0$ (where $T$ is the thermal energy of the dust grains) for an infinite, isotropic lattice when $\alpha = 90^\text{o}$, by using the lattice summation technique. This provides a reliable basis with high accuracy for validation of our MD simulations. The lattice summation was performed by summing the contribution of about $10^9$ neighboring grains on a perfect lattice characterized by the lattice spacing $b$, the aspect ratio $\nu$ and the rhombic angle $\phi$. In addition, in contrast to our previous studies of structural phase transitions in 2D complex plasma composed of ferromagnetic grains with intrinsic magnetic dipole moments \cite{Feldmann} where the density was kept constant, in this case the pressure is kept constant as $\eta$ is varied. The pressure was computed from the diagonal elements of the pressure tensor, which in this case has the form: 
\begin{equation}\label{eq:stress}
p_{\gamma}=\frac{1}{b}\sum_{r_\gamma<0}\frac{r_\gamma}{|{\bf r}|}\nabla_{\bf r}U(r), 
\end{equation} 
where $\gamma$ denotes the Cartesian coordinates ($x$ or $y$), ${\bf r}$ is the distance between the particle at the origin (0,0) and another lattice particle. Summation is performed for particles located on a half-plane. $U(r)$ is the inter-particle pair potential energy, including electrostatic and magnetic contributions. In the calculations the lattice is oriented along the $x$-axes, but due to the force isotropy of a perfect hexagonal lattice, the calculated pressure value does not depend on the lattice orientation. More graphically this is the force per unit length acting on a fictitious line of particles inserted at $x=0$ (or $y=0$) interacting with all particles on one side only. 

\begin{figure}[htb]
\includegraphics[width=0.9\columnwidth]{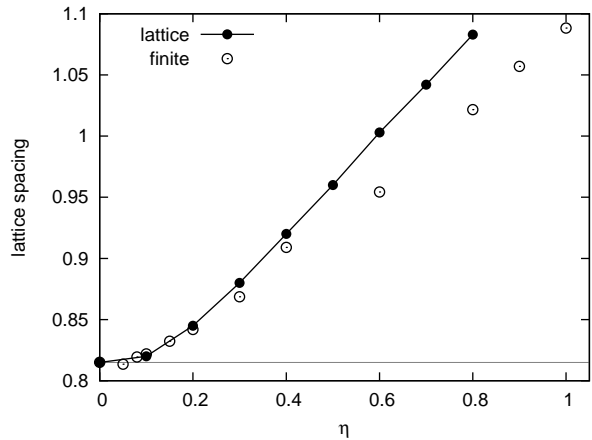}
\caption{\label{fig:11}  
Lattice spacing $b$ (in units of $\lambda_D$) from ``infinite'' lattice summation and from finite MD simulations versus $\eta$, for $\alpha = 90^\text{o}$ ($\nu = 1$, $\phi = 60^\text{o}$).} 
\end{figure}

The configuration with the minimal total energy was sought. The initial lattice parameters were adopted from the $\eta=0$ MD simulation ($\lambda_D \equiv 1$, $\nu=1$, $\phi=60^\text{o}$, $b=0.81$) and the initial pressure value, which was kept constant during the subsequent $\eta>0$ calculation, was evaluated for this initial lattice. The results are shown in Fig.~\ref{fig:11}, which compares the lattice spacing as a function of $\eta$ for the infinite lattice with the MD simulation results for the finite system. Note that in the finite case, the lattice spacing is an average over the central part of the particle cloud. The comparison shows good agreement for the lower magnetization ($\eta < 0.4$) cases, where the deformation of the finite dust cloud is not too large ($\Delta b/b = 10\%$) and confinement can be approximated by the constant pressure condition. 

\section{Possible experimental parameters}

To aid in the design of a possible experimental realization of ground state structures studied in this paper, we estimate a range of  possible plasma and dust parameters necessary to observe such effects.  The quantity $\eta$ is a figure-of-merit, which characterizes the relative strength of the magnetic dipole-dipole to electrostatic interaction between neighboring grains. Assuming that the grain is spherical, with radius $R$, and that it can be characterized by a magnetic permeability $\mu$, its induced magnetic dipole moment can be expressed as \cite{Yaro04}
\begin{equation}\label{eq:7}
M = \frac{\mu_0}{4\pi}R^3 \left( \frac{\mu -1}{\mu + 2} \right) B.
\end{equation}
Expressing the magnitude of the grain charge as $q = R \vert \phi_s\vert$ where $\phi_s$ is the grain surface potential, we have that 
\begin{equation}\label{eq:8}
\eta = \sqrt{\mu_0\varepsilon_0}\frac{M}{q \lambda_D} \sim 0.03\, \frac{R^2(\mu{\rm m}) B(G)}{\phi_s(V) \lambda_D (\mu{\rm m})}\, \left(\frac{\mu -1}{\mu+2} \right).
\end{equation}
For example, consider a plasma with $T_e \sim 2$ eV and $n_i \sim 10^8$ cm$^{-3}$ so that the effective Debye length in the sheath, given approximately by the ion Debye length with $T_i \sim T_e$, is about $\lambda_D \sim 1$ mm. Assuming  that $\mu = 4$, $R = 5$ $\mu$m, $B = 5000$ G, $\phi_s = 2$ V, we obtain $\eta \sim 0.94$. Thus for these dust and plasma parameters, varying the external magnetic field from 0 to 5000 G can vary $\eta$ from 0 to about 1.  Another possibility is a denser plasma, with $T_e \sim 2$ eV and $n_i \sim 10^{10}$ cm$^{-3}$, and the other parameters the same as in the last example. In this case, varying the magnetic field from 0 to 500 G can vary $\eta$ from 0 to about 1. Thus it seems that there could be a range of reasonable experimental parameters for observing the variation of lattice parameters and structures predicted in this paper.

\section{Summary}

The ground state configuration of a 2D dusty plasma crystal composed of super-paramagnetic grains immersed in an external magnetic field has been investigated using MD simulations with parameters that may be close to realizable experimental conditions. Since the magnetic dipole moments of the grains are induced by the external magnetic field, the dipole moments of the grains all lie in the same direction. This study determined the dependence of the lattice parameters and structure on the parameter $\eta$ (which characterizes the relative strength of the magnetic dipole-dipole to electrostatic interactions) and $\alpha$ (the angle between the direction of the magnetic dipole moment and the lattice plane).  It was found that, for a given set of dust and plasma parameters, it may be possible to vary the lattice spacing within a factor of about 2 by changing the magnitude of the external magnetic field or the direction of the field with respect to the dust layer. Correspondingly, the particle density can be varied by about a factor of 2. Moreover, the lattice structure can be tuned from triangular (hexagonal) to almost rectangular; depending on $\eta$ and $\alpha$, the rhombic angle can vary between  60 and 80 degrees, and the aspect ratio between 1 and 2. It was shown that there could be sets of reasonable experimental parameters for observing the effects discussed in this paper. There are other interesting dusty plasma physics issues that could be studied. For example, it would be very interesting to see under what conditions the attractive alignment force may be large enough to overcome the possible rotation of the dust cloud due to an ion drag force in the case where the ions are magnetized (see e.g. \cite{Kon}).

\begin{acknowledgments}
This work was partially supported  by NSF Grants PHY 0715227 and PHY 0903808, NASA Grant NNX10AR54G, DOE Grant DE-FG02-04ER54804, Hungarian Grants OTKA-K-77653, OTKA-PD-75113, MTA-NSF/102, and the J\'anos Bolyai Research Foundation of the Hungarian Academy of Sciences.
\end{acknowledgments}

%\bibliography{magref.bib}

%merlin.mbs 2010-03-15 4.21a (PWD, AO, DPC)

\end{document}